\documentclass{ifacconf}

\counterwithin*{section}{part}

\usepackage{graphicx}
\usepackage{natbib}

\usepackage{todonotes}
\usepackage{xcolor}
\usepackage{amsmath,amssymb}
\usepackage{mathtools}
\usepackage{enumerate} 
\usepackage{xcolor}
\usepackage[absolute]{textpos} 

\usepackage[capitalise]{cleveref}

\newtheorem{assumption}{Assumption}
\newtheorem{corollary}{Corollary}
\newtheorem{definition}{Definition}
\newtheorem{lemma}{Lemma}
\newtheorem{proposition}{Proposition}
\newtheorem{remark}{Remark}
\newtheorem{theorem}{Theorem}

\Crefname{alg}{Algorithm}{Algorithms}
\Crefname{assumption}{Assumption}{Assumptions}
\Crefname{definition}{Definition}{Definitions}
\Crefname{proposition}{Proposition}{Propositions}
\Crefname{remark}{Remark}{Remarks}

\DeclareMathOperator*{\argmin}{arg\,min}

\newcommand{\abs}[1]{\lvert #1 \rvert}
\newcommand{\norm}[1]{\| #1 \|}
\newcommand{\overbar}[1]{\mkern 1.5mu\overline{\mkern-1.5mu#1\mkern-1.5mu}\mkern 1.5mu}
\newcommand{\st}{\text{ s.t. }}
\newcommand{\zs}{z_\mathrm{s}}
\newcommand{\vs}{v_\mathrm{s}}
\newcommand{\zn}{z_N^\ast}
\newcommand{\zi}{z_\infty^\ast}
\newcommand{\zntc}{z_N^{\ast\mathrm{tc}}}
\newcommand{\vntc}{v_N^{\ast\mathrm{tc}}}
\newcommand{\znuc}{z_N^{\ast\mathrm{uc}}}
\newcommand{\vnuc}{v_N^{\ast\mathrm{uc}}}
\newcommand{\vn}{v_N^\ast}
\newcommand{\vi}{v_\infty^\ast}

\newcommand{\kappaf}{\bar{\kappa}_\mathrm{f}}
\newcommand{\Lmax}{L_\pi^\mathrm{max}}
\newcommand{\Lpi}{L_\pi}
\newcommand{\Vtc}{V^\mathrm{tc}}
\newcommand{\Vuc}{V^\mathrm{uc}}
\newcommand{\Jcl}{J^\mathrm{cl}}
\newcommand{\Jcltc}{J^\mathrm{cl,tc}}
\newcommand{\Jcluc}{J^\mathrm{cl,uc}}
\newcommand{\N}{\mathbb{N}}
\newcommand{\Ukappa}{\overbar{\mathbb{U}}_\kappa^T}
\newcommand{\R}{\mathbb{R}}

\newcommand{\X}{\overbar{\mathbb{X}}}
\newcommand{\Z}{\overbar{\mathbb{Z}}}
\newcommand{\W}{\mathbb{W}}
\newcommand{\Xf}{\overbar{\mathbb{X}}_\mathrm{f}}
\newcommand{\Xinit}{\overbar{\mathbb{X}}_\mathrm{init}}
\newcommand{\zcl}{z_\mathrm{cl}^+}
\newcommand{\Zcl}{\mathbb{Z}_\mathrm{cl}^+}
\newcommand{\Vf}{V_\mathrm{f}}

\newcommand{\Kinf}{\mathcal{K}_\infty}
\newcommand{\setA}[1]{\{ #1 \}}
\newcommand{\setB}[2]{\{ #1 \, | \, #2 \}}

\begin{document}

\begin{frontmatter}

\title{Transient Performance of Tube-based Robust Economic Model Predictive Control\thanksref{footnoteinfo}} 

\thanks[footnoteinfo]{This work was supported by Deutsche Forschungsgemeinschaft (DFG, German Research Foundation) under Grants AL 316/12-2 and MU 3929/1-2 - 279734922. The authors thank the International Max Planck Research School for Intelligent Systems (IMPRS-IS) for supporting Lukas Schwenkel.}

\author[First]{Christian Klöppelt} 
\author[Second]{Lukas Schwenkel} 
\author[Second]{Frank Allgöwer}
\author[First]{Matthias A. Müller}

\address[First]{Institute of Automatic Control, Leibniz University Hannover, Germany, (e-mail: \{kloeppelt,mueller\}@irt.uni-hannover.de)}
\address[Second]{Institute for Systems Theory and Automatic Control, University of Stuttgart, Germany, (e-mail: \{schwenkel,allgower\}@ist.uni-stuttgart.de)}

\begin{abstract}
    In this paper, we provide non-averaged and transient performance guarantees for recently developed, tube-based robust economic model predictive control (MPC) schemes. In particular, we consider both tube-based MPC schemes with and without terminal conditions. We show that the closed-loop performance obtained by applying such MPC schemes is approximately optimal when evaluated both on finite and infinite time horizons. These performance bounds are similar to those derived previously for nominal economic MPC. The theoretical results are discussed in a numerical example.
\end{abstract}


\end{frontmatter}

\begin{textblock*}{\textwidth}(1.5cm,29.3cm)
	\small{
		© 2021 The Authors. This work has been accepted to IFAC for publication under a Creative Commons Licence CC-BY-NC-ND.\\
		10.1016/j.ifacol.2021.08.520
	}
\end{textblock*}


\section{INTRODUCTION}

In addition to classical stabilizing model predictive control (MPC) methods, economic MPC (EMPC) methods \citep[see][]{Faulwasser2018} have increasingly moved into the focus of the research community in recent years. With these schemes an economic cost criterion, which is possibly not positive definite with respect to a steady state, is directly minimized. In \cite{Amrit2011} EMPC with terminal conditions and later in \cite{Gruene2013} EMPC without terminal conditions were introduced. One of the main results of both papers was to give an upper bound for the closed-loop asymptotic average performance of the EMPC scheme. However, these bounds are such that---in principle---the performance on any finite time interval can be arbitrarily bad, necessitating the development of stronger, non-averaged performance guarantees. Such performance guarantees were established in \cite{Grune2015} for the case with terminal conditions and in \cite{Gruene2014,Gruene2016} for the case without terminal conditions. Namely, in these references it was shown that the closed-loop infinite horizon non-averaged performance is equal to the performance of the infinite horizon optimal control problem (OCP), up to an error term vanishing with an increasing prediction horizon. Moreover, concerning the closed-loop finite horizon non-averaged performance, it could be shown that the closed-loop trajectories resulting from application of EMPC are optimal among all trajectories leading into a neighborhood of the optimal steady state, again up to an error term which vanishes with an increasing prediction horizon. This performance measure is referred to as transient performance or transient optimality, as it gives an estimate of the closed-loop performance during a transient time interval. 

Since real systems are subject to disturbances and model uncertainties, it is desirable to design robust controllers. One possibility for robust MPC is tube-based MPC, where an artificial nominal state is controlled and the real state is guaranteed to lie in a robust positively invariant set around this nominal state \citep[see][]{Mayne2005}. As shown in \cite{Bayer2014}, when using the classic tube-based setups with an economic stage cost, a deterioration in performance may occur. This was remedied in \cite{Bayer2014} by averaging the stage cost over all possible real states and in \cite{Bayer2016} by considering the worst case stage cost. Both variants were transferred in \cite{Schwenkel2019} to the case without terminal conditions. Other tube-based EMPC methods can be found in \cite{Lucia2014,Broomhead2015,Dong2018}. For all these approaches, only estimates for the closed-loop asymptotic averaged performance could be stated until now. On the other hand, estimates for the non-averaged and transient performance of tube-based EMPC methods do not exist up to now to the best of our knowledge.

The main contribution of this paper is to provide estimates for these performance measures. It is shown that the nominal closed-loop sequences resulting from the application of tube-based MPC lead to desirable transient and non-averaged performance results that are similar to those of nominal EMPC. In addition, estimates for the closed-loop performance of the real closed-loop state trajectory can be given. The results are provided for the case with and the case without terminal conditions. Moreover, we provide a numerical example, which clarifies the differences of the transient performance estimates for robust and nominal EMPC.

The paper is structured as follows. In \cref{sec:problem_setup}, the problem setup and the assumptions used later on are described. In \cref{sec:preliminary_results}, preliminary results based on the so-called turnpike property are specified, which are later used to estimate the non-averaged performance. \cref{sec:mpc_with_tc} and \cref{sec:mpc_without_tc} provide the non-averaged and transient performance results for the case with terminal conditions and for the case without terminal conditions, respectively. In \cref{sec:numerical_example} a numerical example is presented. \cref{sec:conclusion} briefly summarizes the results.

{\emph{Notation:} We denote the set of continuous, strictly increasing functions $\alpha : \R_{\geq0}\to\R_{\geq0}$ with $\alpha(0) = 0$ by $\mathcal{K}$ and the set of continuous, strictly decreasing functions $\delta : \R_{\geq0}\to\R_{\geq0}$ with $\delta(t)\to 0$ as $t\to \infty$ by $\mathcal{L}$. Moreover, the subset of unbounded functions in $\mathcal{K}$ is denoted by $\mathcal{K}_\infty$ and the set of functions $\beta:\R_{\geq0}\times\R_{\geq0}\to\R_{\geq0}$ with $\beta(\cdot,t)\in\mathcal{K}$ and $\beta(x,\cdot)\in\mathcal{L}$ for all fixed $x,t\geq 0$ by $\mathcal{KL}$. The Minkowski set addition for sets $A, B \subseteq \R^n$ is denoted by $A\oplus B \coloneqq \setB{x+y\in\R^n}{x\in A,y\in B}$. An $\epsilon$-ball around $x'\in\R^n$ is denoted by $B_\epsilon(x')\coloneqq \setB{x\in\R^n}{\norm{x-x'}\leq \epsilon}$. We denote the cardinality of a set $Q$, i.e., the number of elements in this set, by $\# Q$.}

\section{PROBLEM SETUP}
\label{sec:problem_setup}

In this section the problem setup is formulated. Later on, we will distinguish between MPC problems with and without terminal conditions. For the scheme with terminal conditions we use the setup from \cite{Bayer2018} and for the scheme without terminal conditions the setup from \cite{Schwenkel2019}. The system to be controlled has the form
\begin{equation}
	\label{eq:system}
	x(t+1) = f(x(t),u(t),w(t)), \quad x(0) = x_0.
\end{equation}
Here, $f:\R^n\times\R^m\times\R^q \to \R^n$ is the nonlinear state transition map, $x(t)\in\R^n$ is the state vector, $u(t)\in\R^m$ is the input vector, and $w(t)\in\R^q$ is an external disturbance. The state and input constraint is given by $\left(x(t),u(t)\right)\in\mathbb{Z}\subseteq\R^n\times \R^m$ for all $t\geq 0$. For the external disturbance, we assume that $w(t)\in\mathbb{W}\subseteq\R^q$ for all $t\geq 0$ applies.

\begin{assumption}
    \label{as:set_z_w}
	The constraint sets $\mathbb{Z}$ and $\mathbb{W}$ are compact, convex, and have a nonempty interior. Furthermore, $\mathbb{W}$ contains the origin in its interior.
\end{assumption}

As it is standard in tube-based MPC, the input $u$ is given by a suitably parameterized feedback, i.e., 
\begin{equation}
	\label{eq:feedback_law}
	u(t) = \pi(x(t),v(t))
\end{equation} 
with $\pi:\R^n\times\R^m \to \R^m$. The variable $v(t)\in\R^m$ is an artificial input which is later used as decision variable in the OCP. To simplify the notation, as in \cite{Bayer2018} we introduce the function
\begin{equation*}
	f_\pi(x,v,w) \coloneqq f(x,\pi(x,v),w)
\end{equation*}
and, based on this, a constraint set
\begin{equation*}
	\mathbb{Z}_\pi \coloneqq \setB{(x,v)\in\R^n\times\R^m}{(x,\pi(x,v))\in\mathbb{Z}}.
\end{equation*}
Moreover, we define the nominal dynamics as
\begin{equation}
	\label{eq:nominal_sys}
	z(t+1) = f_\pi(z(t),v(t),0), \quad z(0) = z_0.
\end{equation} 
where $z(t)\in\R^n$ is an artificial nominal state. Finally, the error dynamics are defined as
\begin{equation}
    \label{eq:error_sys}
	e(t+1) = f_\pi(x(t),v(t),w(t)) - f_\pi(z(t),v(t),0)
\end{equation}
and, based on this, a robust control invariant (RCI) set is introduced.

\begin{definition}{\citep[Definition 1]{Bayer2018}}
	A compact set $\Omega \subseteq \R^n$ is robust control invariant (RCI) for the error system \eqref{eq:error_sys} if there exists a feedback law \eqref{eq:feedback_law} such that for all $x,z \in \R^n$ and $v\in\mathbb{R}^m$ with $e = x-z \in \Omega$ and $(x,v) \in \mathbb{Z}_\pi$ and for all $w\in \mathbb{W}$ it holds that $e^+ = f_\pi(x,v,w)-f_\pi(z,v,0) \in\Omega$.
\end{definition}

The concept of tube-based MPC is based on the idea that, by the use of a control law parameterized by \eqref{eq:feedback_law}, the actual state $x(t)$ is guaranteed to stay in an RCI set $\Omega$ around the nominal state $z(t)$ for all $t\geq 0$. The RCI set can now be used to define the tightened constraints for the nominal state $z(t)$ and the input $v(t)$ as 
\begin{equation*}
	\Z \coloneqq \setB{(z,v)\in\R^n\times\R^m}{(z+\epsilon,v)\in\mathbb{Z}_\pi\ \forall \epsilon \in\Omega}.
\end{equation*}
In general, the computation of RCI sets for nonlinear systems is a non-trivial task. For further information on how to compute RCI sets for nonlinear and linear systems see 
\cite{Bayer2018} and the references therein. For the rest of this paper it is assumed that an RCI set exists.

\begin{assumption}
    \label{as:rci}
	For system \eqref{eq:system} there exists an RCI set $\Omega$ such that $\Omega$ and $\Z$ have a nonempty interior.
\end{assumption}

The goal of EMPC is to minimize a continuous stage cost $L : \R^n\times\R^m \to \R$, which may not be positive definite with respect to a steady state. For further simplification the notation
\begin{equation*}
	L_\pi(x(t), v(t)) \coloneqq L(x(t), \pi (x(t),v(t)))
\end{equation*}
is introduced. As it was shown in \cite{Bayer2014}, the performance of the tube-based EMPC may be improved by including information about the disturbance in the stage cost. This is done, e.g., by using the stage cost
\begin{equation*}
	L_\pi^\mathrm{int} (z,v) \coloneqq \int_\Omega L_\pi(z+\epsilon, v) \ \mathrm{d}\epsilon
\end{equation*}
in which an average over all possible real states within the RCI set $\Omega$ is taken. Alternatively, it is possible to use the worst case stage cost
\begin{equation*}
	L_\pi^\mathrm{max} (z,v) \coloneqq \max_{\epsilon\in\Omega} L_\pi(z+\epsilon,v)
\end{equation*}
as it was shown in \cite{Bayer2016}. For simplicity, $\ell$ is used in the following as a general placeholder for the stage costs $L_\pi$, $L_\pi^\mathrm{int}$ or $L_\pi^\mathrm{max}$. In addition, it is assumed that the stage cost is Lipschitz continuous.

\begin{assumption}
    \label{as:lipschitz}
	The stage cost $\ell: \mathbb{Z}_\pi \to \R$ is Lipschitz continuous on $\mathbb{Z}_\pi$ with Lipschitz constant $\kappa_\ell>0$.
\end{assumption}

With this stage cost the nominal OCP can now be formulated:
\begin{subequations}
    \label{eq:ocp1}
	\begin{alignat}{3}
		V_N(z_0) = & \min_{z(\cdot),v(\cdot)} & \quad & J_N(z_0,v(\cdot)) + \Vf(z(N)) \\
		& \ \ \, \st && z(t+1) = f_\pi(z(t),v(t),0), \label{eq:nb1} \\
		&&& (z(t),v(t)) \in \Z, \label{eq:nb2} \\ 
		&&& \forall \, t\in \setA{0 \dots N-1}, \nonumber \\ 
		&&& z(0) = z_0, \label{eq:nb3} \\ 
		&&& z(N) \in \Xf,
	\end{alignat}
\end{subequations}
with 
\begin{equation}
	\label{eq:cost}
	J_N(z_0,v(\cdot))= \sum_{t=0}^{N-1} \ell( z(t),v(t) ).
\end{equation}
The minimizers for the OCP with the prediction horizon $N\in\N$ and the initial nominal state $z_0$ are denoted by $\zn(\cdot;z_0)$ and $\vn(\cdot;z_0)$. We define the infinite horizon OCP with the cost 
\begin{equation}
	\label{eq:infty_cost}
	J_\infty(z_0,v(\cdot)) = \limsup_{N\to\infty} J_N(z_0,v(\cdot))
\end{equation}
as
\begin{equation}
	\label{eq:infty_ocp}
	V_\infty(z_0) = \inf_{z(\cdot),v(\cdot)\;\text{s.t.}\;\text{(\ref{eq:nb1},\ref{eq:nb2},\ref{eq:nb3})}} J_\infty(z_0,v(\cdot)).
\end{equation}
The minimizers for this OCP are denoted by $\zi(\cdot;z_0)$ and $\vi(\cdot;z_0)$. In the following, we assume that the optimal input sequences $v_N^\ast$ exist for all $N\in\N\cup\setA{\infty}$.\footnote{If this assumption does not hold, one can extend \cref{pr:turnpike_property} to turnpike properties for suboptimal solutions as in \cite{Grune2015} and obtain similar results. For simplicity, we omit this technicality and work with optimal solutions.} Note that given the later on introduced \cref{as:reachability,as:controllability,as:dissipativity}, we can wlog. assume  $\abs{V_\infty(z)} < \infty$ by replacing $\ell(z,v)$ with $\ell(z,v) - \ell(\zs,\vs)$. The upper bound follows immediately from $\ell(\zs,\vs)=0$ being reachable in finite time due to \cref{as:reachability,as:controllability}. The lower bound follows from \cref{as:dissipativity} \citep[compare also][Lemma 4.2]{Grune2015}.

As already mentioned, a distinction is made between OCPs with and without terminal conditions. For the case with terminal conditions, the terminal cost $\Vf: \R^n\to \R$ and the terminal region $\Xf\subseteq\R^n$ are introduced, on which we impose the following assumption which is standard in economic MPC \citep[compare, e.g.,][]{Amrit2011,Faulwasser2018}:

\begin{assumption}{\citep[Assumption 2]{Bayer2016}}
    \label{as:terminal}
	The terminal region $\Xf \subseteq \R^n$ is closed and contains $\zs$ in its interior. There exists a local auxiliary controller $\kappaf: \Xf \to \R^m$ such that for all $z \in \Xf$ it holds that
	\begin{enumerate}[(i)]
		\item $\left(z, \kappaf(z) \right) \in \Z$,
		\item $f_\pi(z,\kappaf (z),0) \in \Xf$, and
		\item $\Vf\left(f_\pi(z,\kappaf (z),0)\right) - \Vf(z) \leq -\ell(z,\kappaf (z)) + \ell (\zs,\vs)$.
	\end{enumerate}
	Moreover, the terminal cost $\Vf$ is continuous on the terminal region $\Xf$.
\end{assumption}

The case without terminal conditions is simply established from \eqref{eq:ocp1} by using $\Vf = 0$ for the terminal cost and $\Xf = \R^n$ for the terminal region. To distinguish between the two cases, the optimal value function and the minimizers of the OCP are denoted by $V_N^\mathrm{tc}(z)$, $\zntc$ and $\vntc$ for the case with terminal conditions and by $V_N^\mathrm{uc}(z)$, $\znuc$ and $\vnuc$ for the unconstrained case, i.e., without terminal conditions. In the following, if certain properties apply for both types of MPC schemes the superscripts are omitted.

In order to set up the tube-based OCP for the real state $x(t)\in\R^n$ measured at time $t$, the nominal initial state is used as a decision variable:
\begin{subequations}
    \label{eq:ocp2}
	\begin{alignat}{3}
		& \min_{z_0} & \quad & V_N(z_0) \\
		& \text{ s.t.} && x(t) \in \setA{z_0} \oplus \Omega.
	\end{alignat}
\end{subequations}
Therefore, we obtain the minimizer $z_0^\ast(x(t))$ and denote the open-loop state and input trajectory at time $t$ as $\zn(\cdot|t) \coloneqq \zn(\cdot;z_0^\ast(x(t)))$ and $\vn(\cdot|t)\coloneqq \vn(\cdot;z_0^\ast(x(t)))$, respectively. Using the initial condition as an optimization variable is standard in many tube-based MPC schemes, compare, e.g., \cite{Mayne2005,Bayer2018}. However, this additional degree of freedom results in closed-loop sequences $\zn(0|t)$, $t\geq 0$ which are not necessarily trajectories of the nominal system \eqref{eq:nominal_sys}. To describe these closed-loop sequences, we introduce the set of all possible next nominal states
\begin{equation*}
	\Zcl(z) \coloneqq \setB{z_0^\ast(x_\mathrm{cl}^+)}{x_\mathrm{cl}^+ \in \setA{f_\pi(z,\vn (0;z),0)}\oplus \Omega}.
\end{equation*}

Now, we define the set of all $z$ such that OCP \eqref{eq:ocp1} with terminal conditions is feasible over the horizon $N$ as
\begin{equation*}
    \begin{split}
        \X_N \coloneqq \setB{z\in\R^n}{& \exists\,(\tilde{z},\tilde{v}):\N\to\Z \st \tilde{z}(0) = z,\,\\ 
        &\tilde{z}(t+1) = f_\pi(\tilde{z}(t),\tilde{v}(t),0),\, \\
        & \forall t\in\setA{0\dots N-1},\, \tilde{z}(N)\in \Xf}
    \end{split}
\end{equation*}
and the set of all $z$ for which there exists an admissible infinite horizon input sequence as
\begin{equation*}
    \begin{split}
        \X_\infty \coloneqq \setB{z\in\R^n}{& \exists\,(\tilde{z},\tilde{v}):\N\to\Z \st \tilde{z}(0) = z,\,\\
        & \tilde{z}(t+1) = f_\pi(\tilde{z}(t),\tilde{v}(t),0),\,\forall t\geq 0}.
    \end{split}
\end{equation*}
For simplicity, in the following $\Xinit$ is used as a general placeholder for $\X_N$ and $\X_\infty$ if properties apply for both cases (tc and uc) of MPC.
We are now in a position to define the economic MPC algorithm as follows:

\begin{alg}
    \label{al:mpc}
	For all $t\geq 0$, given $x(t)$, solve \eqref{eq:ocp2} to obtain $\vn(\cdot|t)$. Afterwards, apply $u(t) = \pi(x(t),\vn(0|t))$ to the real system \eqref{eq:system}.
\end{alg}

Moreover, we will make use of the concept of the so called robust optimal steady state \citep[ROSS, cf.][]{Bayer2018}, which is the steady state minimizing the stage cost within the constraint set:
\begin{equation*}
	(\zs,\vs) \coloneqq \argmin_{(z,v)\in\Z,z=f_\pi(z,v,0)} \ell(z,v).
\end{equation*}

Regarding the ROSS the following assumptions are made.

\begin{assumption}
    \label{as:ross}
	The ROSS lies in the interior of the constraint set. 
\end{assumption}

The next assumption is standard in economic MPC. It ensures that the nominal system is optimally operated at steady state and closed-loop convergence can be ensured for suitable nominal MPC algorithms, compare \cite{Faulwasser2018}.

\begin{assumption}
    \label{as:dissipativity}
	The nominal system \eqref{eq:nominal_sys} is strictly dissipative on $\Z$ with respect to the supply rate $s(z,v) = \ell(z,v) - \ell(\zs,\vs)$, i.e. there is $\alpha_\ell \in \Kinf$ and a bounded storage function $\lambda : \R^n \to \R$ such that for all $(z,v)\in\Z$ it holds
	\begin{equation*}
		\lambda(f_\pi(z,v,0)) - \lambda(z) \leq s(z,v) - \alpha_\ell(\|(z,v) - (\zs,\vs)\|).
	\end{equation*}
\end{assumption}

\begin{assumption}{\cite[Assumption 8]{Schwenkel2019}}
    \label{as:strong_dissipativity}
	Let \cref{as:dissipativity} hold. For all $z\in\Xinit$ and all $\zcl \in \Zcl(z)$ the following dissipation inequality holds
	\begin{equation*}
        \begin{split}
            \lambda(\zcl) - \lambda(z) \leq &  s(z,\vn(0;z)) \\
            & - \alpha_\ell(\|(z,\vn(0;z)) - (\zs,\vs)\|).
        \end{split}
	\end{equation*}
\end{assumption}

\cref{as:strong_dissipativity} is a stronger version of \cref{as:dissipativity}, which ensures convergence of the closed-loop (nominal) state sequence $\zn(0|t)$, $t\geq 0$, to the ROSS. For more information we refer to \cite[Remark 2]{Schwenkel2019}. Furthermore, we assume exponential reachability \citep[cf.][Assumption 4.2]{Faulwasser2018}, local $M$-step controllability \cite[cf.][Section 3.7]{Sontag2013} of the ROSS, and a continuity condition on the optimal value function. These assumptions are standard in economic MPC without terminal constraints \cite{Gruene2013,Gruene2014,Faulwasser2018}.

\begin{assumption}
    \label{as:reachability}
	The ROSS $(\zs,\vs)$ of the nominal system \eqref{eq:nominal_sys} is exponentially reachable.
\end{assumption}

\begin{assumption}
    \label{as:controllability}
    The nominal system \eqref{eq:nominal_sys} is locally $M$-step controllable at $(\zs,\vs)$.
\end{assumption}

\begin{assumption}
	\label{as:cont_ovf}
	There exists $\alpha_V \in \Kinf$ such that for all $N\in\N$ and all $z \in \X$ it holds
	\begin{equation*}
		\abs{V_N(z) - V_N(\zs)} \leq \alpha_V(\norm{z-\zs}).
	\end{equation*}
\end{assumption}

This last assumption implies that $V_N$ is continuous at the ROSS, it holds for suitable choices of the terminal cost and constraints \citep[compare][Assumption 3.6]{Grune2015}, and can also be satisfied if no terminal conditions are used \citep[cf.][Theorem 6.4]{Gruene2013}.

\section{PRELIMINARY RESULTS}
\label{sec:preliminary_results}

In this section, we state some preliminary results. All results described here can be applied to the case with as well as to the case without terminal conditions, which is why the superscripts $\mathrm {tc}$ and $\mathrm {uc}$ are omitted. The first proposition is the turnpike property. It states that if the horizon $N$ is long enough, the optimal trajectories of the OCP \eqref{eq:ocp1} are "most of the time" near the ROSS. 

\begin{proposition}
	\label{pr:turnpike_property}
	Let \cref{as:set_z_w,as:rci,as:lipschitz,as:ross,as:reachability,as:dissipativity} hold. Then there exists $c < \infty$ such that for all $\epsilon > 0$ and all $z\in \Xinit$ and the following statements hold:
	For all $N\in\N\cup\setA{\infty}$ the inequality
			\begin{equation}
				\label{eq:cardinality}
				\#\mathbb{Q}_\epsilon (N,z) \leq \frac{c}{\alpha_\ell(\epsilon)}
			\end{equation}
			holds with 
			\begin{equation*}
					\begin{split}
							\mathbb{Q}_\epsilon(N,z) = \setB{ & k\in\setA{0,\dots,N-1}}{ \\
							& \norm{(\zn(k;z),\vn(k;z)) - (\zs,\vs)}\geq \epsilon}.
					\end{split}
			\end{equation*}
\end{proposition}
\begin{pf}
	The proof for the finite horizon case without terminal conditions can be found in \cite[Proposition 4.1]{Faulwasser2018}. The proofs for the infinite horizon case and the case with terminal conditions are analogous. 
	\hfill $\qed$
\end{pf}

Now the turnpike property can be used to prove the following lemma.

\begin{lemma}
	\label{le:lemma1_ls}
	Consider the finite horizon OCP \eqref{eq:ocp1} and the infinite horizon OCP \eqref{eq:infty_ocp}. Let \cref{as:set_z_w,as:rci,as:lipschitz,as:ross,as:reachability,as:controllability,as:dissipativity} be satisfied. Then there exists a $\rho>0$ for which the following statements hold:
	\begin{enumerate}[(i)]
		\item For all $y,z\in\Xinit$, all $\delta\in\left(0,\rho\right]$, and all $P\in\N$ with $P,\dots , P+M \not\in \mathbb{Q}_\delta (N,z)\cup\mathbb{Q}_\delta (N,y)$ it holds
		\begin{equation*}
			\| V_{N-P}(z_N^{\ast} (P;y)) - V_{N-P}(z_N^{\ast} (P;z)) \| \leq \delta.
		\end{equation*}
		\item For all $y,z,\in\Xinit$, all $\delta\in\left(0,\rho\right]$, and all $P\in\N$ with $P,\dots , P+M \not\in \mathbb{Q}_\delta (\infty,z)\cup\mathbb{Q}_\delta (\infty,y)$ it holds
		\begin{equation*}
			\| V_{\infty}(z_\infty^{\ast} (P;y)) - V_{\infty}(z_\infty^{\ast} (P;z)) \| \leq \delta.
		\end{equation*}
		\item For all $z\in\Xinit$, all $\delta\in\left(0,\rho\right]$, and all $P \in\N$ with $P,\dots , P+M \not\in  \mathbb{Q}_\delta (N,z)\cup\mathbb{Q}_\delta (\infty,z)$ it holds
		\begin{equation*}
			\| J_P(z,v_N^{\ast} (\cdot;z)) - J_P(z, v_\infty^{\ast} (\cdot;z)) \| \leq \delta.
		\end{equation*}
	\end{enumerate}
	The sets $\mathbb{Q}_\delta(N,z)$ and $\mathbb{Q}_\delta(\infty,z)$ are defined analogously to \cref{pr:turnpike_property}.
\end{lemma}
\begin{pf}
	The proof to this lemma is analogous to the one of \cite[Lemma 1]{Schwenkel2019}. The only difference lies in (ii) and (iii), where the infinite horizon case is considered, and in the two additional \cref{as:dissipativity,as:reachability}, which are needed to ensure that the infinite horizon cost is finite. To prove these parts, we simply replace the finite horizon dynamic programming principle by its infinite horizon counterpart. \hfill $\qed$
\end{pf}

This lemma states that, due to the turnpike property, OCPs with the same prediction horizon but different initial values have end pieces with similar costs (part (i) and (ii)) and OCPs with the same initial values but one with finite and one with infinite prediction horizon have start pieces with similar costs (part (iii)). In the next lemma we exploit this property to show that the decrease of each finite horizon optimal cost is a lower bound to the decrease of the infinite horizon optimal cost. This is our key technical lemma which will be exploited in the following sections in order to establish non-averaged and transient performance bounds.

\begin{lemma}
	\label{lemma_diff_infty}
    Let \cref{as:set_z_w,as:rci,as:lipschitz,as:ross,as:reachability,as:controllability,as:dissipativity} hold. Then there exists $\delta_1\in\mathcal{L}$ such that the inequality
    \begin{equation}
        \label{eq:V_diff_infty}
    	V_N(y) - V_N(z) \leq V_\infty(y) - V_\infty(z) + \delta_1(N)
    \end{equation}
    holds for all sufficiently large $N\in\N$ and for all $z,y\in\Xinit$.
\end{lemma}
\begin{pf}
	Note that it suffices to show \eqref{eq:V_diff_infty} for sufficiently large $N$, since it can always be satisfied for small $N$ by choosing $\delta_1$ large enough due to boundedness of $V_N$ and $V_\infty$. We consider two finite and two infinite horizon OCPs, one of each with the initial condition $z$ and $y$, respectively. In order to apply \cref{le:lemma1_ls} to all of these four OCPs there has to be a $P\in\setA{0, \dots , N-1}$ such that
	\begin{equation}
	\label{eq:set_q_prime}
		P, \dots , P+M \not\in \mathbb{Q}' 
	\end{equation}
	holds with $\delta = \delta(N)$, and    
	\begin{equation*}
		\mathbb{Q}' \coloneqq \mathbb{Q}_{\delta}(N,z)\cup \mathbb{Q}_{\delta}(\infty,z)\cup\mathbb{Q}_{\delta}(N,y)\cup\mathbb{Q}_{\delta}(\infty,y).
	\end{equation*}
	To this end, we choose
	\begin{equation*}
		\delta(N) \coloneqq \alpha_\ell^{-1}\left(\frac{4c(M+1)}{N-M-1}\right).
	\end{equation*}
	It follows from \eqref{eq:cardinality} that
	\begin{equation*}
		\# \mathbb{Q}_{\delta (N)}(N,\xi) \leq \frac{N-M-1}{4(M+1)}
	\end{equation*}
	and
	\begin{equation*}
		\# \mathbb{Q}_{\delta (N)}(\infty,\xi) \leq \frac{N-M-1}{4(M+1)}
	\end{equation*}
	for $\xi=z,y$, and thus,
	\begin{equation*}
		\# \mathbb{Q}' \leq \frac{N}{M+1} - 1.
	\end{equation*}
	This implies that there exists at least one $P\in\setA{0,...,N-1}$ such that \eqref{eq:set_q_prime} holds. If we now choose $N$ sufficiently large such that $\delta(N)\leq\rho$, we can apply \cref{le:lemma1_ls}, which yields
	\begin{align}
		\begin{split}
			V_{N}(z) - V_{N}(y) & = J_P(z,\vn(\cdot;z)) - J_P(y,\vn(\cdot;y)) \nonumber \\
			& \quad + V_{N-P}(\zn(P;z)) - V_{N-P}(\zn(P;y)) 
		\end{split} \\
		\begin{split}
			& \leq J_P(z,\vn(\cdot;z)) - J_P(y,\vn(\cdot;y)) \nonumber \\
			& \quad + \delta(N) 
		\end{split} \\
		\begin{split}
			& \leq J_P(z,\vi(\cdot;z)) - J_P(y,\vi(\cdot;y)) \label{eq:ineq2} \\
			& \quad + 3\delta(N) 
		\end{split}
	\end{align}
	where the first equality follows from the dynamic programming principle, the first inequality holds due to \cref{le:lemma1_ls} (i) and the second due to (iii). We can now use the dynamic programming principle
	\begin{equation*}
		V_\infty(\xi) = J_P(\xi,v_\infty^\ast(\cdot;\xi)) + V_\infty(\zi(P;\xi))
	\end{equation*}
	for $\xi=z$ and $\xi=y$, which again follows from the dynamic programming principle. Together with \eqref{eq:ineq2}, this yields
	\begin{align*}
		\begin{split}
			V_{N}(z) - V_{N}(y) & \leq V_\infty(z) - V_\infty(y) - V_\infty(\zi(P;z)) \\
			& \quad + V_\infty(\zi(P;y)) + 3\delta(N)
		\end{split} \\
			& \leq V_\infty(z) - V_\infty(y) + 4\delta(N),
	\end{align*}
	where the last inequality holds due to \cref{le:lemma1_ls} (ii).	Finally, we obtain \eqref{eq:V_diff_infty} with $\delta_1(N) = 4\delta(N)$. \hfill $\qed$
\end{pf}

\section{PERFORMANCE RESULTS WITH TERMINAL CONDITIONS}
\label{sec:mpc_with_tc}

This section contains our main results for the tube-based EMPC with terminal conditions, i.e., non-averaged and transient performance bounds. Therefore, we denote the finite horizon performance for the closed-loop nominal sequences starting at $\zntc(0|0)\in\X_N$ up to a time step $T\in\N$ as
\begin{equation}
    \label{eq:finite_cl_performance_tc}
	\Jcltc_T(\zntc(0|0)) \coloneqq \sum_{t=0}^{T-1} \ell(\zntc(0|t),\vntc(0|t)),
\end{equation}
and the infinite horizon closed-loop performance as
\begin{equation}
	\label{eq:infinite_cl_performance_tc}
    \Jcltc_\infty(\zntc(0|0)) \coloneqq \limsup_{T\to\infty} \Jcltc_T(\zntc(0|0)).
\end{equation}

Firstly, we recall the convergence results \citep[cf.][Theorem 4.17]{Bayer2017} for the setup with terminal conditions.

\begin{proposition}
	\label{pr:stability_tc}
	Let \cref{as:set_z_w,as:rci,as:lipschitz,as:terminal,as:ross,as:dissipativity,as:strong_dissipativity} hold. Then there exists a $\beta\in\mathcal{KL}$ with
	\begin{equation}
		\label{eq:convergence_tc}
		\norm{\zntc(0|t) - \zs} \leq \beta \left(\norm{\zntc(0|0) - \zs}, t \right) 
	\end{equation}
	for all closed-loop sequences $\zntc(0|t+1)\in \Zcl(\zntc(0|t))$, $t\geq 0$ with $\zntc(0|0)\in\X_N$.
\end{proposition}
\begin{pf}
	See \cite[Theorem 4.17]{Bayer2017}.
\end{pf}

Note, that by standard MPC arguments, OCP \eqref{eq:ocp2} in \cref{al:mpc} is recursively feasible for all initial states $z\in\X_N$ \citep[cf.][Theorem 1]{Bayer2016}.

The first performance measure we introduce is the so called non-averaged performance. The following result states that the non-averaged asymptotic performance of the closed-loop nominal state sequences is equal to the infinite-horizon optimal performance $\Vtc_\infty(z)$ up to an error term vanishing with $N\to\infty$. 

\begin{theorem}
    \label{th:nap_tc}
	Let \cref{as:set_z_w,as:rci,as:lipschitz,as:terminal,as:ross,as:reachability,as:controllability,as:dissipativity,as:strong_dissipativity,as:cont_ovf} hold. Then there is a $\delta_1 \in\mathcal{L}$ such that the inequality
    \begin{equation}
        \label{eq:nap_tc}
        \Jcltc_\infty(\zntc(0|0)) \leq V_\infty(\zntc(0|0)) + \delta_1(N)
    \end{equation}
    holds for all $\zntc(0|0)\in\X_N$ and for all $N \in \N$.
\end{theorem}
\begin{pf}
	Assume $\ell(\zs,\vs) = 0$ without loss of generality. We note again that it suffices to show \eqref{eq:nap_tc} for sufficiently large $N$, since the inequality can be satisfied for small $N$ by choosing $\delta_1$ large enough (due to boundedness of $V_\infty$ and $\Jcltc_\infty$).\footnote{The latter follows from \eqref{eq:boundedness_Jcltc} for $T\to\infty$ together with the fact that according \cref{pr:stability_tc}, $\zntc(0|T)\to \zs$ for $T\to\infty$.} From \cref{as:terminal} and using standard MPC arguments (i.e., using the shifted previously optimal solution appended by the local auxiliary controller from \cref{as:terminal} as a candidate solution), it follows that
	\begin{equation*}
		\ell(\zntc(0|0),\vntc(0|0)) \leq \Vtc_N(\zntc(0|t)) - \Vtc_N(\zntc(0|t+1))
	\end{equation*}
	holds \cite[compare][]{Amrit2011}. Summing up this inequality from $t=0$ up to $t=T-1$ and using the fact that the right-hand side results in a telescoping sum yields for the closed-loop cost
	\begin{equation}
		\label{eq:boundedness_Jcltc}
		\Jcltc_T(\zntc(0|0)) \leq \Vtc_N(\zntc(0|0)) - \Vtc_N(\zntc(0|T))
	\end{equation}
	Due to recursive feasibility $\zntc(0|0),\zntc(0|T)\in\X_N$. Therefore, \cref{lemma_diff_infty} can be applied for sufficiently large $N$. This leads to
	\begin{equation}
		\label{eq:nap_t}
		\Jcltc_T(\zntc(0|0)) \leq V_\infty(\zntc(0|0)) - V_\infty(\zntc(0|T)) + \delta_1(N)
	\end{equation}
	For $T\to \infty$, \cref{pr:stability_tc} yields $\zntc(0|T)\to\zs$. Furthermore, using standard economic MPC arguments involving the so-called rotated cost function and assuming without loss of generality that $\Vf(\zs)=0$, it follows that $\Vtc_N(z_s)=0$, compare, e.g., \cite{Amrit2011,Grune2015}. Hence, by \cref{as:cont_ovf}, we obtain $V_\infty(\zntc(0|T))\to 0$, and thus, \eqref{eq:nap_tc} holds. \hfill $\qed$
\end{pf}

\begin{remark}	
    It should be noted that the results for non-averaged performance determined here are similar to those for nominal EMPC from \cite{Grune2015}. The only difference between the two results stems from the fact that in \eqref{eq:nap_tc} the left-hand side considers the infinite horizon performance of the closed-loop state sequence, which does not have to satisfy the nominal dynamics \eqref{eq:nominal_sys}, while on the right-hand side we have state trajectories, which satisfy these dynamics.
\end{remark}

Next, we state measures for the transient performance of the closed-loop system. For nominal EMPC, transient performance results were established in \cite{Grune2015}. Similar to the nominal case, we define the transient performance as the performance of the closed-loop nominal state sequences, which start at $\zntc(0|0)$ at time $t=0$ and end up in a ball $B_\kappa(\zs)$ around the ROSS at time $T$. To this end, we define $\kappa = \beta \left(\norm{\zntc(0|0) - \zs}, T \right)$ with $\beta\in\mathcal{KL}$ from \eqref{eq:convergence_tc}. To compare the closed-loop behavior with other trajectories, we define the set of all admissible input sequences which steer the nominal state from $z$ to $B_\kappa(\zs)$ in time $T$ as
\begin{equation*}
	\overbar{\mathbb{U}}_{B_\kappa(\zs)}^T(z) \coloneqq \setB{v(\cdot)\in\overbar{\mathbb{U}}^T(z)}{\norm{\phi(T,z,v(\cdot)) - \zs} \leq \kappa}
\end{equation*}
where 
\begin{equation*}
    \begin{split}
    	\overbar{\mathbb{U}}^T(z) \coloneqq \setB{v(\cdot)\in\R^m}{&\exists z(\cdot)\in\R^n \st z(0)= z,\,\\
    	& z(t+1) = f_\pi(z(t),v(t),0), \, \\
    	& (z(t),v(t))\in\Z\,\forall t\in\setA{0\dots T-1}}
    \end{split}
\end{equation*}
denotes the set of all admissible input sequences of length $T$ and $\phi(t,z,v(\cdot))$ denotes the solution to \eqref{eq:nominal_sys} for the input sequence $v(\cdot)$ at time $t$, starting at $z$. Moreover, we introduce the notation $\Ukappa = \overbar{\mathbb{U}}_{B_\kappa(\zs)}^T(\zntc(0|0))$ to simplify the following expressions. The following theorem states that applying the input sequence provided by \cref{al:mpc} yields, approximately, the best closed-loop performance among all input sequences that drive the nominal state to a $\kappa$-neighborhood of $\zs$.
\begin{theorem}
    \label{th:tp_tc}
    Let \cref{as:set_z_w,as:rci,as:lipschitz,as:terminal,as:ross,as:dissipativity,as:strong_dissipativity,as:reachability,as:controllability,as:cont_ovf} hold. Then there are $\delta_1,\delta_2\in\mathcal{L}$ such that the inequality
	\begin{equation}
		\label{eq:tp_tc}
		\begin{split}
		\Jcltc_T(\zntc(0|0)) & \leq \inf_{v(\cdot)\in\Ukappa} J_T(\zntc(0|0),v(\cdot)) \\
		& \quad + \delta_1(N) + \delta_2(T)
		\end{split}
	\end{equation}
	holds with $\kappa = \beta\left(\|\zntc(0|0) - \zs \|,T\right)$ from \eqref{eq:convergence_tc} for all $N,T\in \N$ and for all $\zntc(0|0) \in \X_N$.
\end{theorem}
\begin{pf}
	We note again that it is sufficient to show \eqref{eq:tp_tc} for sufficiently large $N$, since the inequality can be satisfied for small $N$ by choosing $\delta_1$ large enough (due to boundedness of $J_T$ and $\Jcltc_T$). Starting at $\zntc(0|0) \in \X_N$, due to $\vi$ being the minimizer of \eqref{eq:infty_ocp} it holds that
	\begin{align*}
		V_\infty(\zntc(0|0)) & = J_\infty(\zntc(0|0),v_\infty^\ast(\cdot;\zntc(0|0))) \\
		\begin{split}
            & \leq \inf_{v(\cdot)\in\Ukappa} \{ J_T (\zntc(0|0),v(\cdot)) \\
            & \quad + V_\infty(\phi(T,\zntc(0|0),v(\cdot))) \}.			
		\end{split}
	\end{align*}
	Together with \eqref{eq:nap_t} this yields
	\begin{align*}
		\Jcltc_T(\zntc(0|0)) &\leq V_\infty(\zntc(0|0)) - V_\infty(\zntc(0|T)) + \delta_1(N) \\
		\begin{split}
            & \leq \inf_{v(\cdot)\in\Ukappa} \{ J_T (\zntc(0|0),v(\cdot)) \\
            & \quad + V_\infty(\phi(T,\zntc(0|0),v(\cdot))) \} \\
            & \quad - V_\infty(\zntc(0|T)) + \delta_1(N)
		 \end{split}
	\end{align*}
    The definition of $\kappa$ and \cref{as:cont_ovf} finally lead to
    \begin{align*}
        \begin{split}
            \Jcltc_T(\zntc(0|0)) &\leq \inf_{v(\cdot)\in\Ukappa} J_T (\zntc(0|0),v(\cdot)) \\
            & \quad + \delta_1(N) + 2\alpha_V(\kappa) \\        	
            & \leq \inf_{v(\cdot)\in\Ukappa} J_T (\zntc(0|0),v(\cdot)) \\
            & \quad + \delta_1(N) + \delta_2(T)
    	\end{split}
    \end{align*}
    with $\delta_2(T) \coloneqq 2\alpha_V\left(\beta(\max_{y\in\X}{\norm{y-\zs}},T)\right)$ where $\X\coloneqq\setB{z\in\R^n}{\exists v\in\R^m \st (z,v)\in\Z}$. The maximum exists, since $\Z$, and therefore, $\X$ are compact. \hfill $\qed$
\end{pf}

\begin{remark}
    As for the non-averaged performance, the results for the transient performance are similar to the nominal case \cite[cf.][]{Grune2015}. Again, the state sequences on the left-hand side are compared to state trajectories satisfying the nominal dynamics \eqref{eq:nominal_sys} on the right-hand side.
\end{remark}

Finally, the performance measures for the closed-loop nominal sequences can be transferred to performance measures for the real closed-loop state trajectory. Therefore, we define 
\begin{align*}
	\mathcal{J}^\mathrm{cl,tc}_T(x(\cdot),N) & \coloneqq \sum_{t=0}^{T-1}L_\pi(x(t),\vntc(0|t)) \\
	\mathcal{J}^\mathrm{cl,tc}_\infty(x(\cdot),N) & \coloneqq \limsup_{T\to\infty} \mathcal{J}^\mathrm{cl,tc}_T(x(t),N).
\end{align*}
In the following, we show that the closed-loop performance measures from \cref{th:nap_tc,th:tp_tc} also hold for the performance of the real state $x$ if the stage cost $\Lmax$ is used. If $\Lpi$ is used as stage cost a (possibly conservative) error term is added to estimate the performance bounds.

\begin{corollary}
    \label{co:performance_tc}
	Let \cref{as:set_z_w,as:rci,as:lipschitz,as:terminal,as:ross,as:dissipativity,as:strong_dissipativity,as:reachability,as:controllability,as:cont_ovf} hold. Then for all $N,T\in\N$, and for all $x\in\X_N \oplus\Omega$ we obtain the following. For $\ell = \Lmax$ the transient performance satisfies
    \begin{equation}
        \label{eq:real_tp_tc_max}
        \mathcal{J}^\mathrm{cl,tc}_T(x,N) \leq \inf_{v(\cdot)\in\Ukappa } J_T (\zntc(0|0),v(\cdot)) + \delta_1(N) + \delta_2(T),
    \end{equation}
    and the non-averaged performance satisfies
    \begin{equation*}
        \label{eq:real_nap_tc_max}
        \mathcal{J}^\mathrm{cl,tc}_\infty(x,N) \leq V_\infty(\zntc(0|0)) + \delta_1(N).
    \end{equation*}
	For $\ell = \Lpi$ the transient performance satisfies
	\begin{equation}
			\label{eq:real_tp_tc}
        \begin{split}
            \mathcal{J}^\mathrm{cl,tc}_T(x,N) & \leq \inf_{v(\cdot)\in\Ukappa} J_T (\zntc(0|0),v(\cdot)) + \delta_1(N)  \\
            & \quad + \delta_2(T) + T\kappa_{\Lpi}\max_{\epsilon\in\Omega}\norm{\epsilon},        	
        \end{split}
	\end{equation}
	and the non-averaged performance satisfies
	\begin{equation*}
        \begin{split}
            \mathcal{J}^\mathrm{cl,tc}_T(x,N) & \leq V_\infty(\zntc(0|0)) - V_\infty(\zntc(0|T)) + \delta_1(N) \\
            & \quad + T\kappa_{\Lpi}\max_{\epsilon\in\Omega}\norm{\epsilon}.
        \end{split}
	\end{equation*}
\end{corollary}
\begin{pf}
	The proof of this corollary is similar to the one of \cite[Corollary 1]{Schwenkel2019}. For the case of $\ell = \Lmax$, the property
	\begin{equation*}
			\Lpi(x,v) \leq \Lmax(z,v)
	\end{equation*}
	holds for all $(z,v)\in\Z$ and all $x\in\setA{z}\oplus\Omega$. Thus, summing up this inequality over $T$ time steps yields
	\begin{equation*}
		\mathcal{J}^\mathrm{cl,tc}_T(x(\cdot),N) \leq \Jcltc_T(\zntc(0|0))
	\end{equation*}
	and therefore the assertion holds. For the case of $\ell = \Lpi$ the property, that
	\begin{equation*}
			\Lpi(x,v) \leq \Lpi(x,v) + \kappa_{\Lpi}\max_{\epsilon\in\Omega}\norm{\epsilon}
	\end{equation*}
	holds for all $(z,v)\in\Z$ and all $x\in\setA{z}\oplus\Omega$, is exploited analogously. \hfill $\qed$
\end{pf}

\section{PERFORMANCE RESULTS WITHOUT TERMINAL CONDITIONS}
\label{sec:mpc_without_tc}

In this section, non-averaged and transient performance is determined for the case without terminal conditions. As in the previous section, the asymptotic convergence properties \cite[cf.][]{Schwenkel2019} are recalled first.

\begin{proposition}
	\label{pr:stability_uc}
	Let \cref{as:set_z_w,as:rci,as:lipschitz,as:ross,as:reachability,as:controllability,as:dissipativity,as:strong_dissipativity} hold. Then there exist $\beta\in\mathcal{KL}$ and $\epsilon_1 \in\mathcal{L}$ such that 
	\begin{equation}
		\label{eq:convergence_uc}
		\norm{\znuc(0|t) - \zs} \leq \max\{\beta \left(\norm{\znuc(0|0) - \zs}, t \right), \epsilon_1(N)\}
	\end{equation}
	holds for all sufficiently large $N\in\N$ and all $\znuc(0|0)\in \X_\infty$, where $\znuc(0|t+1) \in \Zcl(\znuc(0|t))$ denotes the closed-loop sequence resulting from \cref{al:mpc}.
\end{proposition}
\begin{pf}
	See \cite[Theorem 2]{Schwenkel2019}.
\end{pf}
Moreover we recall the following Lemma.
\begin{lemma}
	Let \cref{as:set_z_w,as:rci,as:lipschitz,as:ross,as:reachability,as:controllability,as:dissipativity} hold. Then there exists $\epsilon_1\in\mathcal{L}$ such that
	\begin{equation}
		\label{eq:perf_ls}
		\ell(z,\vnuc(0;z)) \leq \Vuc_N(z) - \Vuc_N(\zcl) + \ell(\zs,\vs) + \epsilon_1(N)
	\end{equation}
	holds for sufficiently large $N\in\N$, all $z\in\X_\infty$ and all $\zcl\in\Zcl(z)$.
\end{lemma}
\begin{pf}
	See \cite[Theorem 1]{Schwenkel2019}.
\end{pf}
With this we can now state the following result on the non-averaged performance.
\begin{theorem}
	\label{th:nap_uc}
    Let \cref{as:set_z_w,as:rci,as:lipschitz,as:ross,as:reachability,as:controllability,as:dissipativity,as:cont_ovf} hold. Then there is a $\delta_3\in\mathcal{L}$ such that the inequality
    \begin{equation}
        \label{eq:nap_uc}
        \Jcluc_T(\znuc(0|0)) \leq V_\infty(\znuc(0|0)) - V_\infty(\znuc(0|T)) + T\delta_3 (N)
    \end{equation}
    holds for all $\znuc(0|0)\in\X_\infty$ and for all $N,T\in\N$.
\end{theorem}
\begin{pf}
    The proof works similar to the proof of \cref{th:nap_tc}. The only difference is that we sum up \eqref{eq:perf_ls} and obtain
	\begin{equation*}
		\Jcl_T(\znuc(0|0)) = V_N(\znuc(0|0)) - V_N(\znuc(0|T)) + T\epsilon_1 (N).
	\end{equation*}
	If we now choose $N$ large enough we may apply \cref{lemma_diff_infty}, which leads to
    \begin{equation*}
		\begin{split}
            \Jcluc_T(\znuc(0|0)) &\leq V_\infty(\znuc(0|0)) - V_\infty(\znuc(0|T)) \\
            &\quad + \delta_1(N) + T\epsilon_1(N).			
		\end{split}
	\end{equation*}
	Thus, we obtain \eqref{eq:nap_uc} with $\delta_3(N) = \delta_1(N) + \epsilon_1(N)$. \hfill $\qed$
\end{pf}

Again, the results are similar to those of the nominal EMPC in \cite{Gruene2016}. The difference to the case with terminal conditions is that in \eqref{eq:nap_uc}, the error term $\delta_3(N)$ now occurs $T$ times in the performance estimate. 
We can interpret \eqref{eq:nap_uc} as that the closed-loop performance on each finite time interval $\left[0,T\right]$ approximates the infinite-horizon performance on this time interval, up to the error term $T\delta_3(N)$.

For the transient performance estimate, state sequences that end up in a $\kappa$-ball around $\zs$ are now considered. At this point $\kappa= \max\{\beta\left(\|\znuc(0|0) - \zs \|,T\right),\epsilon_1(N)\}$ is defined, since the closed-loop state sequences only converge into a $\epsilon_1$-neighborhood of $\zs$.

\begin{theorem}
	\label{th:tp_uc}
    Let \cref{as:set_z_w,as:rci,as:lipschitz,as:terminal,as:ross,as:dissipativity,as:strong_dissipativity,as:reachability,as:controllability,as:cont_ovf} hold. Then there exist $\delta_3,\delta_4\in\mathcal{L}$ such that the inequality
	\begin{equation*}
        \begin{split}
            \Jcluc_T(\znuc(0|0)) & \leq \inf_{v(\cdot)\in\Ukappa} \Jcluc_T(\znuc(0|0),v(\cdot)) \\
            & \quad + T\delta_3(N) + \delta_4(T)
		\end{split}
	\end{equation*}
	holds with $\kappa = \max\{\beta\left(\|\znuc(0|0) - \zs \|,T\right),\epsilon_1(N)\}$ from \eqref{eq:convergence_uc} for all $N,T\in \N$ and all $\znuc(0|0) \in \X_\infty$.
\end{theorem}
\begin{pf}
	The proof is analogous to the one of \cref{th:tp_tc}.
\end{pf}

Comparing the transient performance measure to the nominal case \cite[cf.][]{Gruene2014}, one notes that these are also similar. Thus, all non-averaged and transient performance estimates for the closed-loop nominal state sequences in the robust case can be analogously stated to the nominal case.

In the last corollary, we again transfer our results to performance measures for the real closed-loop state trajectory.

\begin{corollary}
    \label{co:performance_uc}
	Let \cref{as:set_z_w,as:rci,as:lipschitz,as:terminal,as:ross,as:dissipativity,as:strong_dissipativity,as:reachability,as:controllability,as:cont_ovf} hold. Then for all $N,T\in\N$, and all $x\in\X_\infty\oplus\Omega$ it holds for $\ell = \Lmax$
    \begin{equation*}
        \mathcal{J}^\mathrm{cl,uc}_T(x,N) \leq \inf_{v(\cdot)\in\Ukappa} J_T (\znuc(0|0),v(\cdot)) + T\delta_3(N) + \delta_4(T)
    \end{equation*}
    for the transient performance, and 
    \begin{equation*}
        \mathcal{J}^\mathrm{cl,uc}_T(x,N) \leq V_\infty(\znuc(0|0)) - V_\infty(\znuc(0|T)) + T\delta_3 (N)
    \end{equation*}
	for the non-averaged performance, and for $\ell = \Lpi$
	\begin{equation*}
        \begin{split}
            \mathcal{J}^\mathrm{cl,uc}_T(x,N) & \leq \inf_{v(\cdot)\in\Ukappa} J_T (\znuc(0|0),v(\cdot)) + T\delta_3(N) \\
            & \quad+ \delta_4(T) + T\kappa_{\Lpi}\max_{\epsilon\in\Omega}\norm{\epsilon}       	
        \end{split}
	\end{equation*}
	for the transient performance, and
	\begin{equation}
		\label{eq:real_tp_uc}
        \begin{split}
            \mathcal{J}^\mathrm{cl,uc}_T(x,N) & \leq V_\infty(\znuc(0|0)) - V_\infty(\znuc(0|T)) \\
            & \quad + T\delta_3(N) + T\kappa_{\Lpi}\max_{\epsilon\in\Omega}\norm{\epsilon}
        \end{split}
	\end{equation}
	for the non-averaged performance.
\end{corollary}
\begin{pf}
	The proof is analogously to the one in \cref{co:performance_tc}.
\end{pf}

\section{NUMERICAL EXAMPLE}
\label{sec:numerical_example}

To illustrate the results, we consider the example from \cite{Grune2015} with an additive disturbance $w(t)\in\R$, i.e., we consider the linear and scalar system
\begin{equation}
	x(t+1) = 2x(t) + u(t) + w(t)
\label{eq:lin_ex}
\end{equation}
with $x(0) = 2$ and the stage cost $L(u) = u^2$, where the state and input constraint is given by $\mathbb{Z} = \left[-2,2\right]\times\left[-3,3\right]$ and the additive disturbance is bounded by $\W = \left[-\bar{w},\bar{w}\right]$. To stabilize the error dynamics, we use the pre-stabilizing feedback $u = \pi(x,v) = Kx + v$ with $K = -1.5$. Therefore, we obtain $f_\pi(x,v,w) = 0.5x +  v + w$ and $L_\pi(x,v) = (Kx + v)^2$ which is used as stage cost in the OCP. As RCI set we use the set $\Omega=2\W$ which is the minimal robust positively invariant set for the error dynamics $e(t+1) = 0.5e(t) + w(t)$. The ROSS is $(\zs,\vs) = (0,0)$ with the related cost $L_\pi(\zs,\vs) = 0$. One can verify that the the nominal system $z(t+1) = f_\pi(z,v,0)$ is strictly dissipative with respect to the storage function $\lambda(z) = -\frac{1}{2}z^2$. For the "tc" case we use the terminal equality constraint $\Xf=\{0\}$ and terminal cost $V_f(z) = 0$. Therefore, \cref{as:set_z_w,as:rci,as:lipschitz,as:terminal,as:ross,as:dissipativity,as:reachability,as:controllability,as:cont_ovf} are satisfied. In this example, \cref{as:strong_dissipativity} is not satisfied, however, it is only needed to ensure optimal operation at the ROSS which can also be achieved by using the additional constraint $\lambda(\zn(0|t+1)) \leq\lambda(\zn(1|t))$ \cite[cf.][Remark 2]{Schwenkel2019}.

\begin{figure}
\begin{center}
	\includegraphics[width=8.4cm]{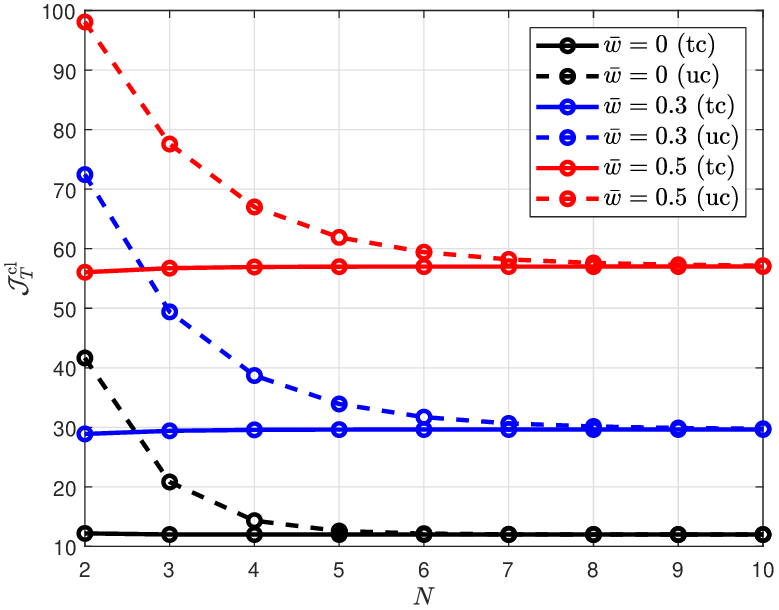}
	\caption{Closed-loop transient performances for $T=20$, $N=2,\dots,10$ and different disturbance bounds (worst-case disturbance sequence)} 
	\label{fig:variable_N}
\end{center}
\end{figure}

In order to show the influence of the different functions $\delta_i$ on the transient performance bounds \eqref{eq:real_tp_tc} and \eqref{eq:real_tp_uc}, we consider the worst-case disturbance sequence ($w(t) = \bar{w} \ \forall t\in\{0,\dots,T\}$) with different bounds $\bar{w}\in\{0,0.3,0.5\}$ and compare the closed-loop transient performances of the real state sequence $\mathcal{J}^\mathrm{cl}_T(x,N)$ for the "uc" case and the "tc" case.  Therefore, similar to the approach in \cite{Grune2015}, we first fix $T=20$ and set $N=2,\dots,10$. The results can be found in \cref{fig:variable_N}, where the solid lines correspond to $\mathcal{J}^\mathrm{cl,tc}_T(x,N)$ and dashed lines to $\mathcal{J}^\mathrm{cl,uc}_T(x,N)$. The black lines restore the nominal (undisturbed) case. The closed-loop transient performance is almost the same for all prediction horizons $N$ when using terminal conditions, i.e. the function $\delta_1(N)$ has only little influence even for small $N$. When omiting terminal conditions, the performance only converges to the "tc" case with a larger $N$. This difference can be explained by the fact that in the "uc" case the closed-loop states only converge to a neighborhood of the ROSS (even without disturbances) and therefore the influence of $\delta_3(N)$ dominates for small $N$ (recall that $\delta_3(N) = \delta_1(N) + \epsilon_1(N)$). A similar behavior can be observed in the disturbed cases with the difference that the performance declines for larger disturbances. This is to be expected and can be explained by the term $T\kappa_{\Lpi}\max_{\epsilon\in\Omega}\norm{\epsilon}$ in \cref{co:performance_tc,co:performance_uc}, which increases with increasing $\bar{w}$, but is independent of $N$.

\begin{figure}
\begin{center}
	\includegraphics[width=8.4cm]{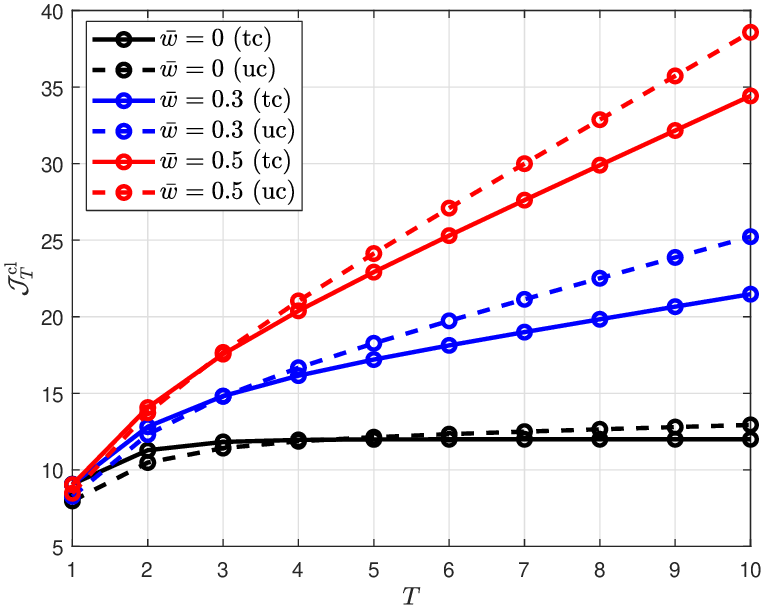}
	\caption{Closed-loop transient performances for $N = 4$ and $T=1,\dots,10$ and different disturbance bounds (worst-case disturbance sequence)} 
	\label{fig:variable_T}
\end{center}
\end{figure}

\cref{fig:variable_T} shows the results for a fixed $N = 4$ and $T=1,\dots,10$. Again, for $\bar{w} = 0$ the undisturbed case restores the results from \cite{Grune2015}. In the "uc" case the term $T\delta_3(N)$ in \cref{co:performance_uc} reduces the performance for increasing $T$ while for the "tc" case the performance converges to a constant (for $\bar{w} = 0)$. With disturbances ($\bar{w}\neq 0$) the performances $\mathcal{J}^\mathrm{cl,tc}_T(x,N)$ and $\mathcal{J}^\mathrm{cl,uc}_T(x,N)$ both increase with increasing $T$. This can be explained by the term $T\kappa_{\Lpi}\max_{\epsilon\in\Omega}\norm{\epsilon}$ in \eqref{eq:real_tp_tc} and \eqref{eq:real_tp_uc}, which yields a steeper slope of the transient performances for larger disturbances.

\section{CONCLUSION}
\label{sec:conclusion}

In this paper, we provided estimates for the closed-loop non-averaged and transient performance of robust economic MPC schemes. In presence of terminal conditions, it was shown that the closed-loop infinite horizon non-averaged performance of the nominal state sequence is approximately equal to the infinite-horizon optimal performance. For the case without terminal conditions, a similar performance estimate was stated on a finite horizon. Moreover, for both cases it was shown that the input sequences computed by the robust EMPC scheme yield, up to an error term, the optimal closed-loop transient performance among all input sequences which drive the nominal state to a neighborhood of the robust optimal steady state. The given results are similar to those of the nominal economic MPC case; the main difference and novelty in the derivation of the bounds in this work is that they hold for closed-loop nominal state sequences, which by the nature of the tube-based MPC scheme do not necessarily satisfy the nominal system dynamics. In addition, the performance estimates could be extended such that upper bounds for the closed-loop performance of the real (disturbed) state could also be provided.

{\footnotesize\bibliography{references}}

\end{document}